\documentclass[aip,apl,reprint,numerical,superscriptaddress,nolongbibliography,groupedaddress]{revtex4-1}
\usepackage{graphicx}
\usepackage{epsfig}
\usepackage{float}

\begin{document}

\title{Real-space imaging of atomic-scale spin textures at nanometer distances}
\author{A. Schlenhoff}
\email[corresp.~author: ]{aschlenh@physnet.uni-hamburg.de}
\homepage[]{http://www.nanoscience.de}
\author{S.~Kovarik}
\thanks{\emph{Present address:} Department of Materials, ETH Zurich, 8093 Zurich, Switzerland}
\author{S. Krause}
\author{R. Wiesendanger}
\affiliation{Department of Physics, University of Hamburg, Jungiusstrasse 11, 20355 Hamburg, Germany}
\begin{abstract}
Spin-polarized scanning tunneling microscopy (SP-STM) experiments on ultrathin films with non-collinear spin textures demonstrate that resonant tunneling allows for atomic-scale spin-sensitive imaging in real space at tip-sample distances of up to $8$\,nm. 
Spin-polarized resonance states evolving between the foremost atom of a magnetic probe tip and the opposed magnetic surface atom are found to provide a loophole from the hitherto existing dilemma of losing spatial resolution when increasing the tip-sample distance in a scanning probe setup. 
Bias-dependent series of SP-STM images recorded via resonant tunneling reveal spin sensitivity at resonance conditions, indicating that the spin-polarized resonance states act as mediators for the spin contrast across the nm-spaced vacuum gap. 
With technically feasible distances in the nm regime, resonant tunneling in SP-STM qualifies for a spin-sensitive read-write technique with ultimate lateral resolution in future spintronic applications.
\end{abstract}
\maketitle

Non-collinear spin textures in ultra-thin film systems are in the focus of ongoing research~\cite{Wiesendanger2016, KCS2017, LMA2020, HBM2011, VHF2018, MPM2019, HLA2018, HGW2017}.
Especially, atomic-scale magnetic skyrmions raise expectations for their application in information technology as logic spin-electronic devices or in recording media~\cite{FRC2017, HKF2017, KW2016, ZEZ2015, ZZF2015, FCS2013, RHM2013}. 
They can be prepared on Si wafers, thereby allowing for large-scale industrial fabrication~\cite{SLF2015}. 
For their characterization and manipulation spin-sensitive techniques with ultimate spatial resolution are required. 
Spin-polarized scanning tunneling microscopy (SP-STM) has successfully demonstrated to be a powerful technique for magnetic imaging in real space with atomic-scale resolution, recording the spatially-resolved spin-polarized tunneling current between the foremost atom of a magnetic probe tip and a magnetic surface~\cite{Wiesendanger2009}.
It is based on the overlap of the tip and the surface wave functions, both evanescent into vacuum~\cite{Wortmann2001,Wiesendanger2009}, and is therefore limited to very small tip-sample distances in the order of a few angstroms. 
These small distances make the technique very fragile in terms of sensitivity to vibrations or for accidental, destructive tip-sample collisions. 
Angstrom distances are also technically challenging for many practical applications in future spin-electronic devices. 
For example, current flying heights of read-write heads in hard disk drives are in the range of a few nanometers. 
In an STM setup, the electron tunneling becomes undetectable when the probe tip and the sample are separated by nanometers. 
A more rigid way to image the surface topography is realized by the so-called topografiner. 
Here, a field-emitted current between a tip and a sample is used~\cite{YWS1971,YWS1972}. 
The spatial resolution is determined by the spot size of the electron beam, that increases with increasing tip-sample distance, being of the size of $3$\,nm for typical tip-sample separations of $(3-5)$\,nm~\cite{SG1994}. 
Spin-polarized resonance states (sp-RSs) are a discrete series of unoccupied spin-split electronic states in the vacuum gap between a tip and a magnetic sample~\cite{KBW2007, Schlenhoff2012, SKK2019, HWB2007, EP1978, HE2016, Crampin2005, NCI1993,BFF1985, BGS1985}.
These states exhibit the same local spin quantization axis as the spin texture of the underlying sample surface~\cite{SKK2019}. 
In an SP-STM setup, the sp-RSs can be addressed by spin-polarized electrons that tunnel resonantly from the magnetic tip via a sp-RS into the surface, resulting in a magnetic image contrast governed by the spin-polarized electron tunneling into the sp-RS~\cite{SKK2019}.

Here, we use SP-STM in the resonant tunneling mode as a magnetic imaging technique at increased tip-sample distance on the magnetic nanoskyrmion lattice of the monolayer (ML) Fe/Ir(111)~\cite{HBM2011} and the spin spiral ground state of the double layer (DL) Fe/Ir(111)~\cite{Hsu2016}. 
We find that tunneling into the sp-RSs allows for resolving the atomic-scale spin textures in real space at tip-sample distances of up to $8$\,nm. 
The magnetic contrast is maximized at resonance conditions, indicating the relevance of the sp-RS as mediator for spin information across a nm-spaced vacuum gap. 

The experiments were performed under ultra-high vacuum conditions with a pressure below $1\cdot10^{-8}$\,Pa using a home-built SP-STM at variable temperatures. 
Within the experimental setup, the entire microscope including the tip was cooled to maximize the thermal stability. 
Antiferromagnetic bulk Cr tips were used as scanning probes~\cite{SKH2010}. 
The Ir(111) substrate was prepared by sputtering with $\rm{Ar}^{+}$ ions at room temperature, followed by annealing under oxygen atmosphere and a high temperature flash. 
Fe was deposited onto Ir(111) by molecular beam epitaxy at elevated substrate temperature. 
\begin{figure}[tb]
\begin{center}
\includegraphics[width=\columnwidth]{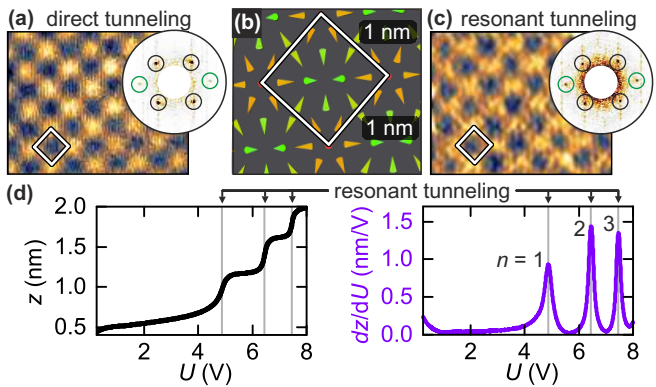}
\caption{\textbf{Imaging the magnetic nanoskyrmion lattice via spin-polarized electron tunneling.}
\textbf{(a)}~Conventional SP-STM constant current image, recorded with electrons tunneling directly into the surface ($U=0.25$\,V), revealing the nanoskyrmion lattice. 
\textbf{(b)}~Schematic spin configuration of the nanoskyrmion lattice. Cones represent surface spin magnetic moments. Magnetic unit cell of (1x1)\,${\rm nm}^{2}$ is indicated. 
\textbf{(c)}~SP-STM constant current image, recorded via resonant electron tunneling ($U=4.2$\,V). 
\textbf{Insets:} Corresponding FFT spectra, revealing spots due to the magnetic square lattice (black circles) and spots (green circles) due to the TAMR effect. 
\textbf{(d)}~Tip-sample displacement $z(U)$ (left) and $dz/dU(U)$ (right). 
The steps in $z(U)$ and peaks in $dz/dU(U)$ indicate resonant electron tunneling.
($I=2$\,nA, $T=25.4$\,K.)
}
\label{fig1}
\end{center}
\end{figure}

The magnetic nanoskyrmion lattice in the ML Fe/Ir(111) is a chiral non-collinear spin texture, stabilized by interfacial Dzyaloshinskii-Moriya interactions, with a square unit cell of $(1$x$1)~\rm{nm}^2$.~\cite{HBM2011} 
In Fig.~\ref{fig1}(a) a SP-STM constant current image of the nanoskyrmion lattice is shown, recorded with an out-of-plane sensitive magnetic tip. 
It reveals a square magnetic contrast of bright and dark areas indicating the spin directions pointing out of and into the surface plane, according to the underlying surface spin texture~\cite{HBM2011} that is schematically shown in Fig.~\ref{fig1}(b). 
The image in Fig.~\ref{fig1}(a) was recorded with electrons tunneling directly into the surface at a tip-sample distance of approx. $0.4$~nm. 
The corresponding two-dimensional (2D) fast Fourier transformation (FFT) spectrum reveals four spots, indicating the reciprocal lattice vectors of the magnetic unit cell of the nanoskyrmion lattice via the tunneling magneto-resistance (TMR) effect~\cite{HBM2011}, and two additional spots arising from the tunneling anisotropic magnetoresistance (TAMR) effect~\cite{HBM2011}. 
They are attributed to a purely electronic effect caused by spin-orbit interaction. 
In Fig.~\ref{fig1}(c), the same area as in (a) was recorded with electrons tunneling into the first sp-RS. 
In order to identify the energy positions of the sp-RSs, scanning tunneling spectroscopy was performed on the sample surface~\cite{BFF1985,BGS1985}. 
Here, a feedback control unit regulates on a constant current $I$ between the probe tip and the sample by adjusting the tip height $z$ while ramping $U$. 
For e$U$ being larger than the sample work function, $I$ is predominantly governed by Fowler-Nordheim field emission~\cite{CPC1999}. 
Consequently, the feedback loop regulates on an approximately constant electric field at the tip, resulting in a tip retraction when increasing $U$. 
In Fig.~\ref{fig1}(d) the tip-sample distance $z(U)$ and its numerical derivative $dz/dU(U)$ are shown. 
Starting at a tip-sample distance of $0.4$\,nm for recording the image shown in Fig.~\ref{fig1}(a), the tip is retracted to a distance of $2.0$\,nm when increasing the sample bias to $8$\,V while keeping $I$ constant. 
For $U$ corresponding to a sp-RS energy, an extra transmission channel opens in addition to the field-emission process via resonant electron tunneling, resulting in steps in $z(U)$ and local maxima in $dz/dU(U)$. 
Note that spin-dependent effects are not visible in Fig.~\ref{fig1}(d), since they are too small to be resolved on these scales~\cite{SKK2019,Schlenhoff2012}.
The image in Fig.~\ref{fig1}(c) also reveals the square magnetic contrast of the nano-skyrmion lattice. 
Accordingly, four spots arising from the magnetic unit cell of the nanoskyrmion lattice via the TMR effect are observable in the corresponding 2D FFT spectrum. 
Additionally, two TAMR spots are observed, indicating a TAMR effect for resonant tunneling conditions~\cite{SOM}. 

\begin{figure}[b]
\begin{center}
\includegraphics[width=\columnwidth]{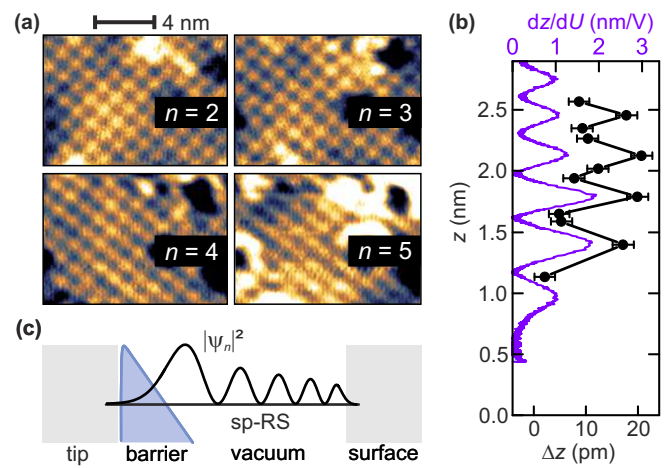}
\caption{\textbf{High-order sp-RSs as mediator for spin contrast at nanometer distances.}
\textbf{(a)}~SP-STM constant current images taken with electrons tunneling into sp-RS $n=2$ ($U=6.5$\,V), $n=3$ ($7.5$\,V), $n=4$ ($8.1$\,V) and $n=5$ ($9.1$\,V).
\textbf{(b)}~Maxima in $dz/dU(z)$ reveal the first $5$ sp-RSs and the tip-sample distance $z$ when tunneling into the respective state. 
Magnetic lattice corrugation $\Delta z$ as a function of tip-sample distance~$z$. 
\textbf{(c)}~Physical picture of the sp-RS electronic wave function spanning the vacuum gap between the tip and the sample. 
 ($I=2$\,nA, $T=25.4$\,K.)
}
\label{fig2}
\end{center}
\end{figure}
To study the bias-dependence, a series of SP-STM constant current images has been performed, recorded in the resonant tunneling mode. 
Exemplary images for state order $n=2$ to $5$ at increasing tip-sample distance are shown in Fig.~\ref{fig2}(a). 
The magnetic nanoskyrmion lattice is resolved for tunneling into all the sp-RSs. 
The peak positions in $dz/dU(z)$, shown in Fig.~\ref{fig2}(b), indicate the tip-sample distances $z$ for tunneling into the respective sp-RS. 
Note, that the $5^{\rm th}$ sp-RS is located about $2.5$\,nm away from the surface. 
In order to analyze the magnetic image contrast, a 2D FFT spectrum has been taken from every image of the series and the spot intensity corresponding to the reciprocal magnetic lattice has been determined as a measure of the magnetic corrugation $\Delta z$. 
It is shown as a function of tip-sample distance $z$ in Fig.~\ref{fig2}(b). 
Whenever in resonance condition, $\Delta z$ is on the order of $20$\,pm. 
The observed lattice corrugation is largest on top of the sp-RS peaks.
The dependency of $\Delta z$ on the resonant tunneling conditions clearly indicates the relevance of the sp-RSs for the magnetic imaging process. 
The lattice corrugation recorded in the direct tunneling mode is $95$\,pm.
Hence, imaging the nanoskyrmion lattice at nm distances decreases the magnetic contrast intensity by about 80\%, compared to direct tunneling. 
However, no loss of magnetic corrugation is found when going to higher-order sp-RSs, as shown in Fig.~\ref{fig2}(b). 
As indicated in Fig.~\ref{fig2}(c), the wave function of a sp-RS spans the distance between the tip and the surface. 
When in resonance, spin-polarized electrons tunnel from the tip into a sp-RS, the latter with a finite probability density at the tip as well as at the surface. 
By this simultaneous overlap of the sp-RS with tip and surface states, the information of the local surface spin orientation is transmitted via the tunneling process. 
Consequently, the sp-RS act as mediator for the spin contrast across the vacuum gap, allowing to perform SP-STM at nanometer distances. 

\begin{figure}[tb]
\begin{center}
\includegraphics[width=\columnwidth]{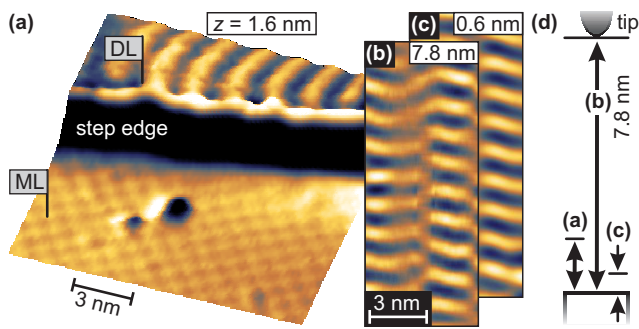}
\caption{\textbf{Real-space imaging of atomic-scale spin textures at nanometer distances.}
\textbf{(a)}~SP-STM constant current image (slightly diff.) of the combined ML and DL Fe/Ir(111) recorded at a tip-sample distance of $z=1.6$\,nm, revealing the ML nanoskyrmion lattice and the DL spin spiral ($U=6.5$\,V, $I=2$\,nA, $T=25.4$\,K). 
Magnetic map of the DL spin spiral, recorded at a tip-sample distance of \textbf{(b)}~$z=7.8$\,nm ($U=18$\,V) and \textbf{(c)}~$z=0.6$\,nm ($U=250$\,mV). 
($I=1$\,nA, $T=30$\,K.)
\textbf{(d)}~Schematic illustration of the tip-sample distances used in (a-c).
}
\label{fig3}
\end{center}
\end{figure}
In Fig.~\ref{fig3}(a), an SP-STM image of the combined ML and DL Fe/Ir(111) is shown, recorded by resonant electron tunneling at a tip-sample distance of $1.6$\,nm. 
On the ML, the magnetic nano-skyrmion lattice is again visible. 
On the DL, a periodic pattern is observable, that is known to arise from a spin spiral ground state~\cite{Hsu2016}. 
Compared to the direct tunneling of spin-polarized electrons, the tip-sample distance is significantly increased by more than one nanometer. 
A magnetic map of the DL spin spiral recorded at a tip-sample distance of $7.8$\,nm is shown in Fig.~\ref{fig3}(b). 
For comparison, a corresponding magnetic map recorded with SP-STM at a distance of $0.6$\,nm is shown in Fig.~\ref{fig3}(c). 
For the tip-sample separation of $7.8$\,nm the magnetic pattern is clearly observed. 
Consequently, even at these large tip-sample distances the sp-RSs act as mediators for the spin contrast.
The relation between the recording distances of the images shown in Fig.~\ref{fig3}(a-c) are illustrated in Fig.~\ref{fig3}(d). 
Note, that direct tunneling at these nm distances is impossible since $I$ decreases by about one order of
magnitude per angstrom when retracting the tip~\cite{Wiesendanger2009}. 
Hence, the tunnel current from the tip into the surface drops by a factor of approx. $10^{-72}$ when retracting the tip by $7.2$\,nm, which is by far below the detection limit of the transimpedance amplifier. 
The resonant tunneling conditions, however, allow for a significant spin-polarized tunnel current that adds to the field emission current, even at large tip-sample distances.

\begin{figure}[b]
\begin{center}
\includegraphics[width=\columnwidth]{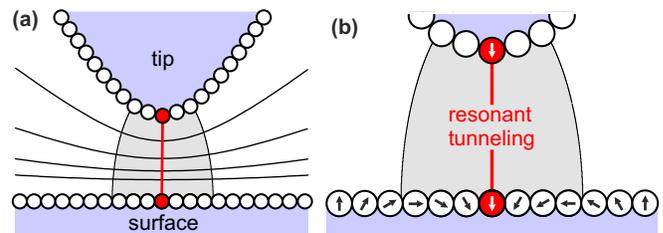}
\caption{\textbf{Field-emission vs. resonant tunneling.}
\textbf{(a)}~Schematic sketch of the equipotential lines between tip and sample, the field emission current and the resonant tunneling current. 
The field-emission current is spatially distributed (light grey region). The resonant tunnel current is confined to the line between tip and sample (indicated in red).
\textbf{(b)}~Closer view. Whereas field emission laterally averages over many surface spins, resonant tunneling conditions are just fulfilled for the atom located directly below the tip apex.
}
\label{fig4}
\end{center}
\end{figure}

The fundamental differences between the topografiner imaging technique and resonant tunneling into sp-RSs is schematically shown in Fig.~\ref{fig4}. 
The topografiner is based on the field-emission current between a tip and a sample.
Here, the spatial resolution deteriorates with increasing tip radius and tip-sample separation~\cite{SG1994}. 
For example, a tip-sample distance of $7.8$\,nm results in a spatial resolution of about $5.5$\,nm at best, thereby averaging over many spin lattice sites on the surface. 
In contrast, for resonant tunneling, the tip in front of a surface spatially confines the resonance condition for the sp-RS to the atomic-scale, being fulfilled only between the foremost atom of the tip and the opposed surface atom, as illustrated in Fig.~\ref{fig4}. 
Note that the resonant tunneling current is independent on the overall shape and radius of the probe tip, allowing to resolve atomic-scale surface spin structures even at nm distances,  irrespective of the actual magnetic probe tip geometry.

In summary, our SP-STM experiments demonstrate resonant spin-polarized tunneling for real-space imaging of atomic-scale spin textures at nm distances. 
The sp-RSs spanning the distance between the magnetic probe tip and the magnetic surface are found to act as mediators for the magnetic image contrast across the nm-spaced vacuum gap. 
The tip-sample distances are in the range of present flying heights of read-write heads in data storage devices. 
In combination with thermally-assisted spin-transfer torque switching under resonant conditions~\cite{Schlenhoff2012},
this spin-sensitive read-write technique may open a pathway towards future technical applications. 
\begin{acknowledgments}
Financial support from the Deutsche Forschungsgemeinschaft via Grant No. SCHL2096/1-2, SCHL2096/1-3 and SPP 2137 "Skyrmionics" is gratefully acknowledged.
\end{acknowledgments}

\bibliographystyle{aipnum4-1}

\begin{thebibliography}{35}%
\makeatletter
\providecommand \@ifxundefined [1]{%
 \@ifx{#1\undefined}
}%
\providecommand \@ifnum [1]{%
 \ifnum #1\expandafter \@firstoftwo
 \else \expandafter \@secondoftwo
 \fi
}%
\providecommand \@ifx [1]{%
 \ifx #1\expandafter \@firstoftwo
 \else \expandafter \@secondoftwo
 \fi
}%
\providecommand \natexlab [1]{#1}%
\providecommand \enquote  [1]{``#1''}%
\providecommand \bibnamefont  [1]{#1}%
\providecommand \bibfnamefont [1]{#1}%
\providecommand \citenamefont [1]{#1}%
\providecommand \href@noop [0]{\@secondoftwo}%
\providecommand \href [0]{\begingroup \@sanitize@url \@href}%
\providecommand \@href[1]{\@@startlink{#1}\@@href}%
\providecommand \@@href[1]{\endgroup#1\@@endlink}%
\providecommand \@sanitize@url [0]{\catcode `\\12\catcode `\$12\catcode
  `\&12\catcode `\#12\catcode `\^12\catcode `\_12\catcode `\%12\relax}%
\providecommand \@@startlink[1]{}%
\providecommand \@@endlink[0]{}%
\providecommand \url  [0]{\begingroup\@sanitize@url \@url }%
\providecommand \@url [1]{\endgroup\@href {#1}{\urlprefix }}%
\providecommand \urlprefix  [0]{URL }%
\providecommand \Eprint [0]{\href }%
\providecommand \doibase [0]{http://dx.doi.org/}%
\providecommand \selectlanguage [0]{\@gobble}%
\providecommand \bibinfo  [0]{\@secondoftwo}%
\providecommand \bibfield  [0]{\@secondoftwo}%
\providecommand \translation [1]{[#1]}%
\providecommand \BibitemOpen [0]{}%
\providecommand \bibitemStop [0]{}%
\providecommand \bibitemNoStop [0]{.\EOS\space}%
\providecommand \EOS [0]{\spacefactor3000\relax}%
\providecommand \BibitemShut  [1]{\csname bibitem#1\endcsname}%
\let\auto@bib@innerbib\@empty
\bibitem [{\citenamefont {Wiesendanger}(2016)}]{Wiesendanger2016}%
  \BibitemOpen
  \bibfield  {author} {\bibinfo {author} {\bibfnamefont {R.}~\bibnamefont
  {Wiesendanger}},\ }\href@noop {} {\bibfield  {journal} {\bibinfo  {journal}
  {Nature Rev. Mater.}\ }\textbf {\bibinfo {volume} {1}},\ \bibinfo {pages}
  {16044} (\bibinfo {year} {2016})}\BibitemShut {NoStop}%
\bibitem [{\citenamefont {Kov{\'a}cs}\ \emph {et~al.}(2017)\citenamefont
  {Kov{\'a}cs}, \citenamefont {Caron}, \citenamefont {Savchenko}, \citenamefont
  {Kiselev}, \citenamefont {Shibata}, \citenamefont {Li}, \citenamefont
  {Kanazawa}, \citenamefont {Tokura}, \citenamefont {Bl{\"u}gel},\ and\
  \citenamefont {Dunin-Borkowski}}]{KCS2017}%
  \BibitemOpen
  \bibfield  {author} {\bibinfo {author} {\bibfnamefont {A.}~\bibnamefont
  {Kov{\'a}cs}}, \bibinfo {author} {\bibfnamefont {J.}~\bibnamefont {Caron}},
  \bibinfo {author} {\bibfnamefont {A.~S.}\ \bibnamefont {Savchenko}}, \bibinfo
  {author} {\bibfnamefont {N.~S.}\ \bibnamefont {Kiselev}}, \bibinfo {author}
  {\bibfnamefont {K.}~\bibnamefont {Shibata}}, \bibinfo {author} {\bibfnamefont
  {Z.-A.}\ \bibnamefont {Li}}, \bibinfo {author} {\bibfnamefont
  {N.}~\bibnamefont {Kanazawa}}, \bibinfo {author} {\bibfnamefont
  {Y.}~\bibnamefont {Tokura}}, \bibinfo {author} {\bibfnamefont
  {S.}~\bibnamefont {Bl{\"u}gel}}, \ and\ \bibinfo {author} {\bibfnamefont
  {R.~E.}\ \bibnamefont {Dunin-Borkowski}},\ }\href@noop {} {\bibfield
  {journal} {\bibinfo  {journal} {Appl. Phys. Lett.}\ }\textbf {\bibinfo
  {volume} {111}},\ \bibinfo {pages} {192410} (\bibinfo {year}
  {2017})}\BibitemShut {NoStop}%
\bibitem [{\citenamefont {Legrand}\ \emph {et~al.}(2020)\citenamefont
  {Legrand}, \citenamefont {Maccariello}, \citenamefont {Ajejas}, \citenamefont
  {Collin}, \citenamefont {Vecchiola}, \citenamefont {Bouzehouane},
  \citenamefont {Reyren}, \citenamefont {Cros},\ and\ \citenamefont
  {Fert}}]{LMA2020}%
  \BibitemOpen
  \bibfield  {author} {\bibinfo {author} {\bibfnamefont {W.}~\bibnamefont
  {Legrand}}, \bibinfo {author} {\bibfnamefont {D.}~\bibnamefont
  {Maccariello}}, \bibinfo {author} {\bibfnamefont {F.}~\bibnamefont {Ajejas}},
  \bibinfo {author} {\bibfnamefont {S.}~\bibnamefont {Collin}}, \bibinfo
  {author} {\bibfnamefont {A.}~\bibnamefont {Vecchiola}}, \bibinfo {author}
  {\bibfnamefont {K.}~\bibnamefont {Bouzehouane}}, \bibinfo {author}
  {\bibfnamefont {N.}~\bibnamefont {Reyren}}, \bibinfo {author} {\bibfnamefont
  {V.}~\bibnamefont {Cros}}, \ and\ \bibinfo {author} {\bibfnamefont
  {A.}~\bibnamefont {Fert}},\ }\href
  {https://doi.org/10.1038/s41563-019-0468-3} {\bibfield  {journal} {\bibinfo
  {journal} {Nature Materials}\ }\textbf {\bibinfo {volume} {19}},\ \bibinfo
  {pages} {34} (\bibinfo {year} {2020})}\BibitemShut {NoStop}%
\bibitem [{\citenamefont {Heinze}\ \emph {et~al.}(2011)\citenamefont {Heinze},
  \citenamefont {von Bergmann}, \citenamefont {Menzel}, \citenamefont {Brede},
  \citenamefont {Kubetzka}, \citenamefont {Wiesendanger}, \citenamefont
  {Bihlmayer},\ and\ \citenamefont {Bl\"ugel}}]{HBM2011}%
  \BibitemOpen
  \bibfield  {author} {\bibinfo {author} {\bibfnamefont {S.}~\bibnamefont
  {Heinze}}, \bibinfo {author} {\bibfnamefont {K.}~\bibnamefont {von
  Bergmann}}, \bibinfo {author} {\bibfnamefont {M.}~\bibnamefont {Menzel}},
  \bibinfo {author} {\bibfnamefont {J.}~\bibnamefont {Brede}}, \bibinfo
  {author} {\bibfnamefont {A.}~\bibnamefont {Kubetzka}}, \bibinfo {author}
  {\bibfnamefont {R.}~\bibnamefont {Wiesendanger}}, \bibinfo {author}
  {\bibfnamefont {G.}~\bibnamefont {Bihlmayer}}, \ and\ \bibinfo {author}
  {\bibfnamefont {S.}~\bibnamefont {Bl\"ugel}},\ }\href@noop {} {\bibfield
  {journal} {\bibinfo  {journal} {Nature Physics}\ }\textbf {\bibinfo {volume}
  {7}},\ \bibinfo {pages} {713} (\bibinfo {year} {2011})}\BibitemShut {NoStop}%
\bibitem [{\citenamefont {McVitie}\ \emph {et~al.}(2018)\citenamefont
  {McVitie}, \citenamefont {Hughes}, \citenamefont {Fallon}, \citenamefont
  {McFadzean}, \citenamefont {McGrouther}, \citenamefont {Krajnak},
  \citenamefont {Legrand}, \citenamefont {Maccariello}, \citenamefont {Collin},
  \citenamefont {Garcia}, \citenamefont {Reyren}, \citenamefont {Cros},
  \citenamefont {Fert}, \citenamefont {Zeissler},\ and\ \citenamefont
  {Marrows}}]{VHF2018}%
  \BibitemOpen
  \bibfield  {author} {\bibinfo {author} {\bibfnamefont {S.}~\bibnamefont
  {McVitie}}, \bibinfo {author} {\bibfnamefont {S.}~\bibnamefont {Hughes}},
  \bibinfo {author} {\bibfnamefont {K.}~\bibnamefont {Fallon}}, \bibinfo
  {author} {\bibfnamefont {S.}~\bibnamefont {McFadzean}}, \bibinfo {author}
  {\bibfnamefont {D.}~\bibnamefont {McGrouther}}, \bibinfo {author}
  {\bibfnamefont {M.}~\bibnamefont {Krajnak}}, \bibinfo {author} {\bibfnamefont
  {W.}~\bibnamefont {Legrand}}, \bibinfo {author} {\bibfnamefont
  {D.}~\bibnamefont {Maccariello}}, \bibinfo {author} {\bibfnamefont
  {S.}~\bibnamefont {Collin}}, \bibinfo {author} {\bibfnamefont
  {K.}~\bibnamefont {Garcia}}, \bibinfo {author} {\bibfnamefont
  {N.}~\bibnamefont {Reyren}}, \bibinfo {author} {\bibfnamefont
  {V.}~\bibnamefont {Cros}}, \bibinfo {author} {\bibfnamefont {A.}~\bibnamefont
  {Fert}}, \bibinfo {author} {\bibfnamefont {K.}~\bibnamefont {Zeissler}}, \
  and\ \bibinfo {author} {\bibfnamefont {C.~H.}\ \bibnamefont {Marrows}},\
  }\href {https://doi.org/10.1038/s41598-018-23799-0} {\bibfield  {journal}
  {\bibinfo  {journal} {Scientific Reports}\ }\textbf {\bibinfo {volume} {8}},\
  \bibinfo {pages} {5703} (\bibinfo {year} {2018})}\BibitemShut {NoStop}%
\bibitem [{\citenamefont {Meyer}\ \emph {et~al.}(2019)\citenamefont {Meyer},
  \citenamefont {Perini}, \citenamefont {von Malottki}, \citenamefont
  {Kubetzka}, \citenamefont {Wiesendanger}, \citenamefont {von Bergmann},\ and\
  \citenamefont {Heinze}}]{MPM2019}%
  \BibitemOpen
  \bibfield  {author} {\bibinfo {author} {\bibfnamefont {S.}~\bibnamefont
  {Meyer}}, \bibinfo {author} {\bibfnamefont {M.}~\bibnamefont {Perini}},
  \bibinfo {author} {\bibfnamefont {S.}~\bibnamefont {von Malottki}}, \bibinfo
  {author} {\bibfnamefont {A.}~\bibnamefont {Kubetzka}}, \bibinfo {author}
  {\bibfnamefont {R.}~\bibnamefont {Wiesendanger}}, \bibinfo {author}
  {\bibfnamefont {K.}~\bibnamefont {von Bergmann}}, \ and\ \bibinfo {author}
  {\bibfnamefont {S.}~\bibnamefont {Heinze}},\ }\href
  {https://doi.org/10.1038/s41467-019-11831-4} {\bibfield  {journal} {\bibinfo
  {journal} {Nature Communications}\ }\textbf {\bibinfo {volume} {10}},\
  \bibinfo {pages} {3823} (\bibinfo {year} {2019})}\BibitemShut {NoStop}%
\bibitem [{\citenamefont {Hsu}\ \emph {et~al.}(2018)\citenamefont {Hsu},
  \citenamefont {R{\'o}zsa}, \citenamefont {Finco}, \citenamefont {Schmidt},
  \citenamefont {Palot{\'a}s}, \citenamefont {Vedmedenko}, \citenamefont
  {Udvardi}, \citenamefont {Szunyogh}, \citenamefont {Kubetzka}, \citenamefont
  {von Bergmann},\ and\ \citenamefont {Wiesendanger}}]{HLA2018}%
  \BibitemOpen
  \bibfield  {author} {\bibinfo {author} {\bibfnamefont {P.-J.}\ \bibnamefont
  {Hsu}}, \bibinfo {author} {\bibfnamefont {L.}~\bibnamefont {R{\'o}zsa}},
  \bibinfo {author} {\bibfnamefont {A.}~\bibnamefont {Finco}}, \bibinfo
  {author} {\bibfnamefont {L.}~\bibnamefont {Schmidt}}, \bibinfo {author}
  {\bibfnamefont {K.}~\bibnamefont {Palot{\'a}s}}, \bibinfo {author}
  {\bibfnamefont {E.}~\bibnamefont {Vedmedenko}}, \bibinfo {author}
  {\bibfnamefont {L.}~\bibnamefont {Udvardi}}, \bibinfo {author} {\bibfnamefont
  {L.}~\bibnamefont {Szunyogh}}, \bibinfo {author} {\bibfnamefont
  {A.}~\bibnamefont {Kubetzka}}, \bibinfo {author} {\bibfnamefont
  {K.}~\bibnamefont {von Bergmann}}, \ and\ \bibinfo {author} {\bibfnamefont
  {R.}~\bibnamefont {Wiesendanger}},\ }\href
  {https://doi.org/10.1038/s41467-018-04015-z} {\bibfield  {journal} {\bibinfo
  {journal} {Nature Communications}\ }\textbf {\bibinfo {volume} {9}},\
  \bibinfo {pages} {1571} (\bibinfo {year} {2018})}\BibitemShut {NoStop}%
\bibitem [{\citenamefont {Hauptmann}\ \emph {et~al.}(2017)\citenamefont
  {Hauptmann}, \citenamefont {Gerritsen}, \citenamefont {Wegner},\ and\
  \citenamefont {Khajetoorians}}]{HGW2017}%
  \BibitemOpen
  \bibfield  {author} {\bibinfo {author} {\bibfnamefont {N.}~\bibnamefont
  {Hauptmann}}, \bibinfo {author} {\bibfnamefont {J.~W.}\ \bibnamefont
  {Gerritsen}}, \bibinfo {author} {\bibfnamefont {D.}~\bibnamefont {Wegner}}, \
  and\ \bibinfo {author} {\bibfnamefont {A.~A.}\ \bibnamefont
  {Khajetoorians}},\ }\href@noop {} {\bibfield  {journal} {\bibinfo  {journal}
  {Nano Lett.}\ }\textbf {\bibinfo {volume} {17}},\ \bibinfo {pages} {5660}
  (\bibinfo {year} {2017})}\BibitemShut {NoStop}%
\bibitem [{\citenamefont {Fert}, \citenamefont {Reyren},\ and\ \citenamefont
  {Cros}(2017)}]{FRC2017}%
  \BibitemOpen
  \bibfield  {author} {\bibinfo {author} {\bibfnamefont {A.}~\bibnamefont
  {Fert}}, \bibinfo {author} {\bibfnamefont {N.}~\bibnamefont {Reyren}}, \ and\
  \bibinfo {author} {\bibfnamefont {V.}~\bibnamefont {Cros}},\ }\href@noop {}
  {\bibfield  {journal} {\bibinfo  {journal} {Nature Rev. Mater.}\ }\textbf
  {\bibinfo {volume} {2}},\ \bibinfo {pages} {17031} (\bibinfo {year}
  {2017})}\BibitemShut {NoStop}%
\bibitem [{\citenamefont {Hsu}\ \emph {et~al.}(2017)\citenamefont {Hsu},
  \citenamefont {Kubetzka}, \citenamefont {Finco}, \citenamefont {Romming},
  \citenamefont {von Bergmann},\ and\ \citenamefont {Wiesendanger}}]{HKF2017}%
  \BibitemOpen
  \bibfield  {author} {\bibinfo {author} {\bibfnamefont {P.-J.}\ \bibnamefont
  {Hsu}}, \bibinfo {author} {\bibfnamefont {A.}~\bibnamefont {Kubetzka}},
  \bibinfo {author} {\bibfnamefont {A.}~\bibnamefont {Finco}}, \bibinfo
  {author} {\bibfnamefont {N.}~\bibnamefont {Romming}}, \bibinfo {author}
  {\bibfnamefont {K.}~\bibnamefont {von Bergmann}}, \ and\ \bibinfo {author}
  {\bibfnamefont {R.}~\bibnamefont {Wiesendanger}},\ }\href
  {https://doi.org/10.1038/nnano.2016.234} {\bibfield  {journal} {\bibinfo
  {journal} {Nature Nanotechnology}\ }\textbf {\bibinfo {volume} {12}},\
  \bibinfo {pages} {123} (\bibinfo {year} {2017})}\BibitemShut {NoStop}%
\bibitem [{\citenamefont {Krause}\ and\ \citenamefont
  {Wiesendanger}(2016)}]{KW2016}%
  \BibitemOpen
  \bibfield  {author} {\bibinfo {author} {\bibfnamefont {S.}~\bibnamefont
  {Krause}}\ and\ \bibinfo {author} {\bibfnamefont {R.}~\bibnamefont
  {Wiesendanger}},\ }\href {\doibase 10.1038/nmat4615} {\bibfield  {journal}
  {\bibinfo  {journal} {Nature Mater.}\ }\textbf {\bibinfo {volume} {15}},\
  \bibinfo {pages} {493} (\bibinfo {year} {2016})}\BibitemShut {NoStop}%
\bibitem [{\citenamefont {Zhang}, \citenamefont {Ezawa},\ and\ \citenamefont
  {Zhou}(2015)}]{ZEZ2015}%
  \BibitemOpen
  \bibfield  {author} {\bibinfo {author} {\bibfnamefont {X.}~\bibnamefont
  {Zhang}}, \bibinfo {author} {\bibfnamefont {M.}~\bibnamefont {Ezawa}}, \ and\
  \bibinfo {author} {\bibfnamefont {Y.}~\bibnamefont {Zhou}},\ }\href
  {https://doi.org/10.1038/srep09400} {\bibfield  {journal} {\bibinfo
  {journal} {Scientific Reports}\ }\textbf {\bibinfo {volume} {5}},\ \bibinfo
  {pages} {9400} (\bibinfo {year} {2015})}\BibitemShut {NoStop}%
\bibitem [{\citenamefont {Zhang}\ \emph {et~al.}(2015)\citenamefont {Zhang},
  \citenamefont {Zhao}, \citenamefont {Fangohr}, \citenamefont {Liu},
  \citenamefont {Xia}, \citenamefont {Xia},\ and\ \citenamefont
  {Morvan}}]{ZZF2015}%
  \BibitemOpen
  \bibfield  {author} {\bibinfo {author} {\bibfnamefont {X.}~\bibnamefont
  {Zhang}}, \bibinfo {author} {\bibfnamefont {G.~P.}\ \bibnamefont {Zhao}},
  \bibinfo {author} {\bibfnamefont {H.}~\bibnamefont {Fangohr}}, \bibinfo
  {author} {\bibfnamefont {J.~P.}\ \bibnamefont {Liu}}, \bibinfo {author}
  {\bibfnamefont {W.~X.}\ \bibnamefont {Xia}}, \bibinfo {author} {\bibfnamefont
  {J.}~\bibnamefont {Xia}}, \ and\ \bibinfo {author} {\bibfnamefont {F.~J.}\
  \bibnamefont {Morvan}},\ }\href {https://doi.org/10.1038/srep07643}
  {\bibfield  {journal} {\bibinfo  {journal} {Scientific Reports}\ }\textbf
  {\bibinfo {volume} {5}},\ \bibinfo {pages} {7643} (\bibinfo {year}
  {2015})}\BibitemShut {NoStop}%
\bibitem [{\citenamefont {Fert}, \citenamefont {Cros},\ and\ \citenamefont
  {Sampaio}(2013)}]{FCS2013}%
  \BibitemOpen
  \bibfield  {author} {\bibinfo {author} {\bibfnamefont {A.}~\bibnamefont
  {Fert}}, \bibinfo {author} {\bibfnamefont {V.}~\bibnamefont {Cros}}, \ and\
  \bibinfo {author} {\bibfnamefont {J.}~\bibnamefont {Sampaio}},\ }\href
  {https://doi.org/10.1038/nnano.2013.29} {\bibfield  {journal} {\bibinfo
  {journal} {Nature Nanotechnology}\ }\textbf {\bibinfo {volume} {8}},\
  \bibinfo {pages} {152} (\bibinfo {year} {2013})}\BibitemShut {NoStop}%
\bibitem [{\citenamefont {Romming}\ \emph {et~al.}(2013)\citenamefont
  {Romming}, \citenamefont {Hanneken}, \citenamefont {Menzel}, \citenamefont
  {Bickel}, \citenamefont {Wolter}, \citenamefont {von Bergmann}, \citenamefont
  {Kubetzka},\ and\ \citenamefont {Wiesendanger}}]{RHM2013}%
  \BibitemOpen
  \bibfield  {author} {\bibinfo {author} {\bibfnamefont {N.}~\bibnamefont
  {Romming}}, \bibinfo {author} {\bibfnamefont {C.}~\bibnamefont {Hanneken}},
  \bibinfo {author} {\bibfnamefont {M.}~\bibnamefont {Menzel}}, \bibinfo
  {author} {\bibfnamefont {J.~E.}\ \bibnamefont {Bickel}}, \bibinfo {author}
  {\bibfnamefont {B.}~\bibnamefont {Wolter}}, \bibinfo {author} {\bibfnamefont
  {K.}~\bibnamefont {von Bergmann}}, \bibinfo {author} {\bibfnamefont
  {A.}~\bibnamefont {Kubetzka}}, \ and\ \bibinfo {author} {\bibfnamefont
  {R.}~\bibnamefont {Wiesendanger}},\ }\href@noop {} {\bibfield  {journal}
  {\bibinfo  {journal} {Science}\ }\textbf {\bibinfo {volume} {341}},\ \bibinfo
  {pages} {636} (\bibinfo {year} {2013})}\BibitemShut {NoStop}%
\bibitem [{\citenamefont {Schlenhoff}\ \emph {et~al.}(2015)\citenamefont
  {Schlenhoff}, \citenamefont {Lindner}, \citenamefont {Friedlein},
  \citenamefont {Krause}, \citenamefont {Wiesendanger}, \citenamefont {Weinl},
  \citenamefont {Schreck},\ and\ \citenamefont {Albrecht}}]{SLF2015}%
  \BibitemOpen
  \bibfield  {author} {\bibinfo {author} {\bibfnamefont {A.}~\bibnamefont
  {Schlenhoff}}, \bibinfo {author} {\bibfnamefont {P.}~\bibnamefont {Lindner}},
  \bibinfo {author} {\bibfnamefont {J.}~\bibnamefont {Friedlein}}, \bibinfo
  {author} {\bibfnamefont {S.}~\bibnamefont {Krause}}, \bibinfo {author}
  {\bibfnamefont {R.}~\bibnamefont {Wiesendanger}}, \bibinfo {author}
  {\bibfnamefont {M.}~\bibnamefont {Weinl}}, \bibinfo {author} {\bibfnamefont
  {M.}~\bibnamefont {Schreck}}, \ and\ \bibinfo {author} {\bibfnamefont
  {M.}~\bibnamefont {Albrecht}},\ }\href@noop {} {\bibfield  {journal}
  {\bibinfo  {journal} {ACS Nano}\ }\textbf {\bibinfo {volume} {9}},\ \bibinfo
  {pages} {5908} (\bibinfo {year} {2015})}\BibitemShut {NoStop}%
\bibitem [{\citenamefont {Wiesendanger}(2009)}]{Wiesendanger2009}%
  \BibitemOpen
  \bibfield  {author} {\bibinfo {author} {\bibfnamefont {R.}~\bibnamefont
  {Wiesendanger}},\ }\href@noop {} {\bibfield  {journal} {\bibinfo  {journal}
  {Rev. Mod. Phys.}\ }\textbf {\bibinfo {volume} {81}},\ \bibinfo {pages}
  {1495} (\bibinfo {year} {2009})}\BibitemShut {NoStop}%
\bibitem [{\citenamefont {Wortmann}\ \emph {et~al.}(2001)\citenamefont
  {Wortmann}, \citenamefont {Heinze}, \citenamefont {Kurz}, \citenamefont
  {Bihlmayer},\ and\ \citenamefont {Bl{\"u}gel}}]{Wortmann2001}%
  \BibitemOpen
  \bibfield  {author} {\bibinfo {author} {\bibfnamefont {D.}~\bibnamefont
  {Wortmann}}, \bibinfo {author} {\bibfnamefont {S.}~\bibnamefont {Heinze}},
  \bibinfo {author} {\bibfnamefont {P.}~\bibnamefont {Kurz}}, \bibinfo {author}
  {\bibfnamefont {G.}~\bibnamefont {Bihlmayer}}, \ and\ \bibinfo {author}
  {\bibfnamefont {S.}~\bibnamefont {Bl{\"u}gel}},\ }\href@noop {} {\bibfield
  {journal} {\bibinfo  {journal} {Phys. Rev. Lett.}\ }\textbf {\bibinfo
  {volume} {86}},\ \bibinfo {pages} {4132} (\bibinfo {year}
  {2001})}\BibitemShut {NoStop}%
\bibitem [{\citenamefont {Young}, \citenamefont {Ward},\ and\ \citenamefont
  {Scire}(1971)}]{YWS1971}%
  \BibitemOpen
  \bibfield  {author} {\bibinfo {author} {\bibfnamefont {R.}~\bibnamefont
  {Young}}, \bibinfo {author} {\bibfnamefont {J.}~\bibnamefont {Ward}}, \ and\
  \bibinfo {author} {\bibfnamefont {F.}~\bibnamefont {Scire}},\ }\href@noop {}
  {\bibfield  {journal} {\bibinfo  {journal} {Phys. Rev. Lett.}\ }\textbf
  {\bibinfo {volume} {27}},\ \bibinfo {pages} {922} (\bibinfo {year}
  {1971})}\BibitemShut {NoStop}%
\bibitem [{\citenamefont {Young}, \citenamefont {Ward},\ and\ \citenamefont
  {Scire}(1972)}]{YWS1972}%
  \BibitemOpen
  \bibfield  {author} {\bibinfo {author} {\bibfnamefont {R.}~\bibnamefont
  {Young}}, \bibinfo {author} {\bibfnamefont {J.}~\bibnamefont {Ward}}, \ and\
  \bibinfo {author} {\bibfnamefont {F.}~\bibnamefont {Scire}},\ }\href@noop {}
  {\bibfield  {journal} {\bibinfo  {journal} {Rev. Sci. Instr.}\ }\textbf
  {\bibinfo {volume} {43}},\ \bibinfo {pages} {999} (\bibinfo {year}
  {1972})}\BibitemShut {NoStop}%
\bibitem [{\citenamefont {S{\'a}enz}\ and\ \citenamefont
  {Garcia}(1994)}]{SG1994}%
  \BibitemOpen
  \bibfield  {author} {\bibinfo {author} {\bibfnamefont {J.~J.}\ \bibnamefont
  {S{\'a}enz}}\ and\ \bibinfo {author} {\bibfnamefont {R.}~\bibnamefont
  {Garcia}},\ }\href@noop {} {\bibfield  {journal} {\bibinfo  {journal} {Appl.
  Phys. Lett.}\ }\textbf {\bibinfo {volume} {65}},\ \bibinfo {pages} {3022}
  (\bibinfo {year} {1994})}\BibitemShut {NoStop}%
\bibitem [{\citenamefont {Kubetzka}, \citenamefont {Bode},\ and\ \citenamefont
  {Wiesendanger}(2007)}]{KBW2007}%
  \BibitemOpen
  \bibfield  {author} {\bibinfo {author} {\bibfnamefont {A.}~\bibnamefont
  {Kubetzka}}, \bibinfo {author} {\bibfnamefont {M.}~\bibnamefont {Bode}}, \
  and\ \bibinfo {author} {\bibfnamefont {R.}~\bibnamefont {Wiesendanger}},\
  }\href@noop {} {\bibfield  {journal} {\bibinfo  {journal} {Appl. Phys.
  Lett.}\ }\textbf {\bibinfo {volume} {91}},\ \bibinfo {pages} {012508}
  (\bibinfo {year} {2007})}\BibitemShut {NoStop}%
\bibitem [{\citenamefont {Schlenhoff}\ \emph {et~al.}(2012)\citenamefont
  {Schlenhoff}, \citenamefont {Krause}, \citenamefont {Sonntag},\ and\
  \citenamefont {Wiesendanger}}]{Schlenhoff2012}%
  \BibitemOpen
  \bibfield  {author} {\bibinfo {author} {\bibfnamefont {A.}~\bibnamefont
  {Schlenhoff}}, \bibinfo {author} {\bibfnamefont {S.}~\bibnamefont {Krause}},
  \bibinfo {author} {\bibfnamefont {A.}~\bibnamefont {Sonntag}}, \ and\
  \bibinfo {author} {\bibfnamefont {R.}~\bibnamefont {Wiesendanger}},\
  }\href@noop {} {\bibfield  {journal} {\bibinfo  {journal} {Phys. Rev. Lett.}\
  }\textbf {\bibinfo {volume} {109}},\ \bibinfo {pages} {097602} (\bibinfo
  {year} {2012})}\BibitemShut {NoStop}%
\bibitem [{\citenamefont {Schlenhoff}\ \emph {et~al.}(2019)\citenamefont
  {Schlenhoff}, \citenamefont {Kovarik}, \citenamefont {Krause},\ and\
  \citenamefont {Wiesendanger}}]{SKK2019}%
  \BibitemOpen
  \bibfield  {author} {\bibinfo {author} {\bibfnamefont {A.}~\bibnamefont
  {Schlenhoff}}, \bibinfo {author} {\bibfnamefont {S.}~\bibnamefont {Kovarik}},
  \bibinfo {author} {\bibfnamefont {S.}~\bibnamefont {Krause}}, \ and\ \bibinfo
  {author} {\bibfnamefont {R.}~\bibnamefont {Wiesendanger}},\ }\href {\doibase
  10.1103/PhysRevLett.123.087202} {\bibfield  {journal} {\bibinfo  {journal}
  {Phys. Rev. Lett.}\ }\textbf {\bibinfo {volume} {123}},\ \bibinfo {pages}
  {087202} (\bibinfo {year} {2019})}\BibitemShut {NoStop}%
\bibitem [{\citenamefont {Hanuschkin}, \citenamefont {Wortmann},\ and\
  \citenamefont {Bl\"ugel}(2007)}]{HWB2007}%
  \BibitemOpen
  \bibfield  {author} {\bibinfo {author} {\bibfnamefont {A.}~\bibnamefont
  {Hanuschkin}}, \bibinfo {author} {\bibfnamefont {D.}~\bibnamefont
  {Wortmann}}, \ and\ \bibinfo {author} {\bibfnamefont {S.}~\bibnamefont
  {Bl\"ugel}},\ }\href {\doibase 10.1103/PhysRevB.76.165417} {\bibfield
  {journal} {\bibinfo  {journal} {Phys. Rev. B}\ }\textbf {\bibinfo {volume}
  {76}},\ \bibinfo {pages} {165417} (\bibinfo {year} {2007})}\BibitemShut
  {NoStop}%
\bibitem [{\citenamefont {Echenique}\ and\ \citenamefont
  {Pendry}(1978)}]{EP1978}%
  \BibitemOpen
  \bibfield  {author} {\bibinfo {author} {\bibfnamefont {P.~M.}\ \bibnamefont
  {Echenique}}\ and\ \bibinfo {author} {\bibfnamefont {J.~B.}\ \bibnamefont
  {Pendry}},\ }\href@noop {} {\bibfield  {journal} {\bibinfo  {journal} {J.
  Phys. C: Solid State Phys.}\ }\textbf {\bibinfo {volume} {11}},\ \bibinfo
  {pages} {2065} (\bibinfo {year} {1978})}\BibitemShut {NoStop}%
\bibitem [{\citenamefont {H{\"o}fer}\ and\ \citenamefont
  {Echenique}(2016)}]{HE2016}%
  \BibitemOpen
  \bibfield  {author} {\bibinfo {author} {\bibfnamefont {U.}~\bibnamefont
  {H{\"o}fer}}\ and\ \bibinfo {author} {\bibfnamefont {P.}~\bibnamefont
  {Echenique}},\ }\href@noop {} {\bibfield  {journal} {\bibinfo  {journal}
  {Surf. Sci.}\ }\textbf {\bibinfo {volume} {643}} (\bibinfo {year}
  {2016})}\BibitemShut {NoStop}%
\bibitem [{\citenamefont {Crampin}(2005)}]{Crampin2005}%
  \BibitemOpen
  \bibfield  {author} {\bibinfo {author} {\bibfnamefont {S.}~\bibnamefont
  {Crampin}},\ }\href@noop {} {\bibfield  {journal} {\bibinfo  {journal} {Phys.
  Rev. Lett.}\ }\textbf {\bibinfo {volume} {95}},\ \bibinfo {pages} {046801}
  (\bibinfo {year} {2005})}\BibitemShut {NoStop}%
\bibitem [{\citenamefont {Nekovee}, \citenamefont {Crampin},\ and\
  \citenamefont {Inglesfield}(1993)}]{NCI1993}%
  \BibitemOpen
  \bibfield  {author} {\bibinfo {author} {\bibfnamefont {M.}~\bibnamefont
  {Nekovee}}, \bibinfo {author} {\bibfnamefont {S.}~\bibnamefont {Crampin}}, \
  and\ \bibinfo {author} {\bibfnamefont {J.~E.}\ \bibnamefont {Inglesfield}},\
  }\href@noop {} {\bibfield  {journal} {\bibinfo  {journal} {Phys. Rev. Lett.}\
  }\textbf {\bibinfo {volume} {70}},\ \bibinfo {pages} {3099} (\bibinfo {year}
  {1993})}\BibitemShut {NoStop}%
\bibitem [{\citenamefont {Binnig}\ \emph {et~al.}(1985)\citenamefont {Binnig},
  \citenamefont {Frank}, \citenamefont {Fuchs}, \citenamefont {Garcia},
  \citenamefont {Reihl}, \citenamefont {Rohrer}, \citenamefont {Salvan},\ and\
  \citenamefont {Williams}}]{BFF1985}%
  \BibitemOpen
  \bibfield  {author} {\bibinfo {author} {\bibfnamefont {G.}~\bibnamefont
  {Binnig}}, \bibinfo {author} {\bibfnamefont {K.~H.}\ \bibnamefont {Frank}},
  \bibinfo {author} {\bibfnamefont {H.}~\bibnamefont {Fuchs}}, \bibinfo
  {author} {\bibfnamefont {N.}~\bibnamefont {Garcia}}, \bibinfo {author}
  {\bibfnamefont {B.}~\bibnamefont {Reihl}}, \bibinfo {author} {\bibfnamefont
  {H.}~\bibnamefont {Rohrer}}, \bibinfo {author} {\bibfnamefont
  {F.}~\bibnamefont {Salvan}}, \ and\ \bibinfo {author} {\bibfnamefont {A.~R.}\
  \bibnamefont {Williams}},\ }\href@noop {} {\bibfield  {journal} {\bibinfo
  {journal} {Phys. Rev. Lett.}\ }\textbf {\bibinfo {volume} {55}},\ \bibinfo
  {pages} {991} (\bibinfo {year} {1985})}\BibitemShut {NoStop}%
\bibitem [{\citenamefont {Becker}, \citenamefont {Golovchenko},\ and\
  \citenamefont {Swartzentruber}(1985)}]{BGS1985}%
  \BibitemOpen
  \bibfield  {author} {\bibinfo {author} {\bibfnamefont {R.~S.}\ \bibnamefont
  {Becker}}, \bibinfo {author} {\bibfnamefont {J.~A.}\ \bibnamefont
  {Golovchenko}}, \ and\ \bibinfo {author} {\bibfnamefont {B.~S.}\ \bibnamefont
  {Swartzentruber}},\ }\href@noop {} {\bibfield  {journal} {\bibinfo  {journal}
  {Phys. Rev. Lett.}\ }\textbf {\bibinfo {volume} {55}},\ \bibinfo {pages}
  {987} (\bibinfo {year} {1985})}\BibitemShut {NoStop}%
\bibitem [{\citenamefont {Hsu}\ \emph {et~al.}(2016)\citenamefont {Hsu},
  \citenamefont {Finco}, \citenamefont {Schmidt}, \citenamefont {Kubetzka},
  \citenamefont {von Bergmann},\ and\ \citenamefont {Wiesendanger}}]{Hsu2016}%
  \BibitemOpen
  \bibfield  {author} {\bibinfo {author} {\bibfnamefont {P.-J.}\ \bibnamefont
  {Hsu}}, \bibinfo {author} {\bibfnamefont {A.}~\bibnamefont {Finco}}, \bibinfo
  {author} {\bibfnamefont {L.}~\bibnamefont {Schmidt}}, \bibinfo {author}
  {\bibfnamefont {A.}~\bibnamefont {Kubetzka}}, \bibinfo {author}
  {\bibfnamefont {K.}~\bibnamefont {von Bergmann}}, \ and\ \bibinfo {author}
  {\bibfnamefont {R.}~\bibnamefont {Wiesendanger}},\ }\href@noop {} {\bibfield
  {journal} {\bibinfo  {journal} {Phys. Rev. Lett.}\ }\textbf {\bibinfo
  {volume} {116}},\ \bibinfo {pages} {017201} (\bibinfo {year}
  {2016})}\BibitemShut {NoStop}%
\bibitem [{\citenamefont {Schlenhoff}\ \emph {et~al.}(2010)\citenamefont
  {Schlenhoff}, \citenamefont {Krause}, \citenamefont {Herzog},\ and\
  \citenamefont {Wiesendanger}}]{SKH2010}%
  \BibitemOpen
  \bibfield  {author} {\bibinfo {author} {\bibfnamefont {A.}~\bibnamefont
  {Schlenhoff}}, \bibinfo {author} {\bibfnamefont {S.}~\bibnamefont {Krause}},
  \bibinfo {author} {\bibfnamefont {G.}~\bibnamefont {Herzog}}, \ and\ \bibinfo
  {author} {\bibfnamefont {R.}~\bibnamefont {Wiesendanger}},\ }\href@noop {}
  {\bibfield  {journal} {\bibinfo  {journal} {Appl. Phys. Lett.}\ }\textbf
  {\bibinfo {volume} {{97}}},\ \bibinfo {pages} {083104} (\bibinfo {year}
  {{2010}})}\BibitemShut {NoStop}%
\bibitem [{\citenamefont {Caama$\tilde{\mathrm{n}}$o}\ \emph
  {et~al.}(1999)\citenamefont {Caama$\tilde{\mathrm{n}}$o}, \citenamefont
  {Pogorelov}, \citenamefont {Custance}, \citenamefont {M$\acute{\rm e}$ndez},
  \citenamefont {Bar$\acute{\rm o}$}, \citenamefont {Veuillen}, \citenamefont
  {G$\acute{\rm o}$mez-Rodr$\acute{\rm i}$guez},\ and\ \citenamefont
  {S$\acute{\rm a}$enz}}]{CPC1999}%
  \BibitemOpen
  \bibfield  {author} {\bibinfo {author} {\bibfnamefont {A.~J.}\ \bibnamefont
  {Caama$\tilde{\mathrm{n}}$o}}, \bibinfo {author} {\bibfnamefont
  {Y.}~\bibnamefont {Pogorelov}}, \bibinfo {author} {\bibfnamefont
  {O.}~\bibnamefont {Custance}}, \bibinfo {author} {\bibfnamefont
  {J.}~\bibnamefont {M$\acute{\rm e}$ndez}}, \bibinfo {author} {\bibfnamefont
  {A.~M.}\ \bibnamefont {Bar$\acute{\rm o}$}}, \bibinfo {author} {\bibfnamefont
  {J.~Y.}\ \bibnamefont {Veuillen}}, \bibinfo {author} {\bibfnamefont {J.~M.}\
  \bibnamefont {G$\acute{\rm o}$mez-Rodr$\acute{\rm i}$guez}}, \ and\ \bibinfo
  {author} {\bibfnamefont {J.~J.}\ \bibnamefont {S$\acute{\rm a}$enz}},\
  }\href@noop {} {\bibfield  {journal} {\bibinfo  {journal} {Surf. Sci.}\
  }\textbf {\bibinfo {volume} {426}},\ \bibinfo {pages} {L420} (\bibinfo {year}
  {1999})}\BibitemShut {NoStop}%
\bibitem [{SOM()}]{SOM}%
  \BibitemOpen
  \href@noop {} {}\bibinfo {howpublished} {see supplemental online materials
  for a detailed discussion of TMR and TAMR observed via resonant
  tunneling.}\BibitemShut {Stop}%
\end{thebibliography}
%

\end{document}